\documentclass[twocolumn,showpacs,preprintnumbers,amsmath,amssymb,superscriptaddress]{revtex4}

\usepackage{graphicx}
\usepackage{dcolumn}
\usepackage{bm}

\let\a=\alpha \let\b=\beta  \let\g=\gamma     \let\d=\delta  \let\e=\varepsilon
  \let\h=\eta    \let\k=\kappa   \let\l=\lambda
\let\m=\mu    \let\n=\nu    \let\x=\xi        \let\p=\pi      \let\r=\rho
\let\s=\sigma \let\t=\tau        
\let\ps=\psi   \let\o=\omega 
 \let\D=\Delta     \let\L=\Lambda  
           
\let\O=\Omega 

\def\\{\hfill\break} \let\==\equiv

\let\io=\infty 

\let\0=\noindent

\def\tende#1{\,\vtop{\ialign{##\crcr\rightarrowfill\crcr
 \noalign{\kern-1pt\nointerlineskip}
 \hskip3.pt${\scriptstyle #1}$\hskip3.pt\crcr}}\,}
\def\otto{\,{\kern-1.truept\leftarrow\kern-5.truept\to\kern-1.truept}\,}

\def\VV{{\cal V}}

\def\T#1{{#1_{\kern-3pt\lower7pt\hbox{$\widetilde{}$}}\kern3pt}}
\def\VVV#1{{\underline #1}_{\kern-3pt
\lower7pt\hbox{$\widetilde{}$}}\kern3pt\,}
\def\W#1{#1_{\kern-3pt\lower7.5pt\hbox{$\widetilde{}$}}\kern2pt\,}

\def\indica{\leaders \hbox to 0.5cm{\hss.\hss}\hfill}
\def\guida{\leaders\hbox to 1em{\hss.\hss}\hfill}
\mathchardef\oo= "0521

\def\xx{{\bf x}}\def\yy{{\bf y}}\def\kk{{\bf k}}
\def\pp{{\bf p}}

\def\be{\begin{equation}}
\def\ee{\end{equation}}
\def\bea{\begin{eqnarray}}\def\eea{\end{eqnarray}}
\def\bp{\begin{pmatrix}}
\def\ep{\end{pmatrix}}
\def\nn{\nonumber}

\def\qed{\raise1pt\hbox{\vrule height5pt width5pt depth0pt}}
\def\bfe{{\bf e}}
\def\pp{{\bf p}}\def\xx{{\bf x}}
 
\def\yy{{\bf y}}\def\kk{{\bf k}}\def\nn{{\bf n}}

\def\bp{{\bar \ps}}

\def\la{{\langle}}               \def\ra{{\rangle}}

\def\nn{\nonumber}
\def\Halmos{\hfill\vrule height10pt width4pt depth2pt \par\hbox to \hsize{}}
\def\pref#1{(\ref{#1})}
\def\virg{\quad,\quad}
\def\qed{\raise1pt\hbox{\vrule height5pt width5pt depth0pt}}

\let\==\equiv

\let\io=\infty
\let\0=\noindent

\begin{document}


\title{Universal Relations for Non Solvable Statistical Models}

\author{G. Benfatto}
\affiliation{Dipartimento di Matematica, Universit\`a di Roma ``Tor Vergata",
00133 Roma, Italy.}
\author{P. Falco}
\affiliation{Mathematics Department, University of British Columbia, Vancouver,
BC Canada, V6T 1Z2}
\author{V. Mastropietro}
\affiliation{Dipartimento di Matematica, Universit\`a di Roma ``Tor Vergata",
00133 Roma, Italy.}

\begin{abstract}
We present the first rigorous derivation of a number of universal relations for
a class of models with continuously varying indices (among which are
interacting planar Ising models, quantum spin chains and 1D Fermi systems), for
which an exact solution is not known, except in a few special cases. Most of
these formulas were conjectured by Luther and Peschel, Kadanoff, Haldane, but
only checked in the special solvable models; one of them, related to the
anisotropic Ashkin-Teller model, is novel.
\end{abstract}

\maketitle

It has long been conjectured, mainly by Kadanoff \cite{K,KB,KW}, Luther and
Peschel \cite{LP} and Haldane \cite{H}, that a number of {\it universal
relations} among critical exponents and other observables hold in a wide
class of models, including planar Ising-like models with quartic interactions,
vertex or Ashkin-Teller models, quantum spin chains and 1D fermionic systems.
Such relations express how the {\it universality principle} works in models
with continuously varying indices: the critical exponents are model dependent
(non-universal) but satisfy model independent formulas, so allowing, for
instance, {\it to express all the exponents in terms of a single one}. The
universal relations have been verified only in certain special exactly solvable
models, but the conjecture is that they are generally valid in a larger class
of models, for which an exact solution is not available.

The interest in this kind of universal relations has been renewed by recent
experiments on materials described by models in this class, like quantum spin
chain models (KCuF3) \cite{L}, carbon nanotubes \cite{Au}, layered structures
\cite{BP} or even 1D Bose systems \cite{IG}. In such systems the critical
exponents depend on the extraordinarily complex and largely unknown microscopic
details of the compounds, but the universal relations allow concrete and
testable predictions for them in terms of a few measurable parameters.

Several attempts in the last thirty years have been devoted to the proof of the
universal relations \cite{PB,N,ZZ,S}, by taking as a starting point the formal
continuum limit (identical for all the models considered here), where extra
Lorentz and Gauge symmetries are verified and make it solvable. Of course,
lattice effects destroy such symmetries and change the exponents; however, this
problem has never been analyzed. On the other hand, not all the relations which
are valid in the special solvable models are generically true; this happens,
for example, for the exponents involved in the dynamic correlations \cite{IG}
and another example will be shown below. It is therefore important to determine
rigorously, and therefore unambiguously, under which conditions and which one
among the relations valid in the solvable models are generically true.

Aim of this letter is to report the first rigorous derivation of several of
such universal relations in a wide class of models, including {\it non
solvable} models; in addition we will also prove a relation which is totally
new.

The simplest class of models in the class of universality we are considering is
coupled Ising models. A configuration $(\s,\s')$ is the product of two
configurations of spins $\s=\{\s_\xx=\pm1\}_{\xx\in \L}$ and
$\s'=\{\s'_\xx=\pm1\}_{\xx\in \L}$. For a finite lattice $\L$, the energy
$H(\s,\s')$ is a function of the parameters $J>0$, $J'>0$ and $\l$
\be\label{11}
\hspace{-.08cm} H =-J\sum_{\xx\in\L\atop j=0,1} \s_{\xx}\s_{\xx+\bfe_j}
-J'\sum_{\xx\in\L\atop j=0,1} \s'_{\xx}\s'_{\xx+\bfe_j} - \l V(\s,\s')
\ee
where $\bfe_0$ and $\bfe_1$ are the horizontal and vertical unit bond.
$V(\s,\s')$ is a quartic interaction, short ranged and symmetric in the
exchange $\s\to \s'$; for instance
\be
V(\s,\s')=\sum_{j=0,1}\sum_{\xx,\yy\in\L} v(\xx-\yy)\s_\xx\s_{\xx+{\bf e}_j}
\s'_\yy\s'_{\yy+{\bf e}_j}\nn
\ee
with $v(\xx)$ a short range potential. It is well known that several models in
Statistical Mechanics can be rewritten as coupled Ising models. In particular
the {\it Ashkin-Teller} model \cite{Ba}, a natural generalization of the Ising
model to four states spins, can be rewritten in the form \pref{11} with
$v(\xx)=\d_{\xx,0}$. Another example is provided by the {\it Eight Vertex}
model, in which $J=J'$ and $V(\s,\s')=
\sum_{j=0,1}\sum_{\xx\in\L}\s_{\xx+j({\bf e}_0+{\bf e}_1)}\s_{\xx+{\bf e}_0}
\s'_{\xx+j({\bf e}_0+{\bf e}_1)}\s'_{\xx+{\bf e}_1}$. An exact solution
\cite{Ba} exists only in the case of the 8V model and {\it not} for the generic
Hamiltonian \pref{11}; even in the case of the 8V model, the correlations have
not been computed and only a few indices can be obtained.

Recently new methods have been introduced in \cite{PS} and \cite{M} to study 2D
statistical mechanics models, which can be considered as a perturbation of the
Ising model. These methods take advantage of the fact that such systems can be
mapped in systems of interacting fermions in $d=1+1$ dimensions. This mapping
was known since a long time \cite{SML}, but only in recent years a great
progress has been achieved in the evaluation of the Grassmann integrals
involved in the analysis of the interacting models, in the context of Quantum
Field Theory and Solid State Physics, so that one can take advantage of this
new technology to get information about statistical mechanics models. At the
moment, when an exact solution is lacking, this is the only way to get rigorous
quantitative information on the properties of such systems. The algorithm is
based on multiscale analysis and allows us to prove convergence of several
thermodynamic functions and correlations up to the critical temperature;
essential ingredients of the analysis are compensations due to the
anticommutativity of Grassmann variables and asymptotic Ward Identities (WI).

By using such methods, it has been proved in \cite{M}, in the case $J=J'$ and
$\l$ small, that the model is critical in the thermodynamic limit at the
inverse temperature $\b_c=T_c^{-1}= \arctan (\sqrt{2}-1)/J +O(\l)$; for $T$
near $T_c$, the specific heat behaves as
\be\label{h10}
C_v\sim \a^{-1} [|T-T_c|^{-\a}-1]
\ee
with $\a$ a continuous non trivial function of $\l$. Moreover, if
$G^\e(\xx-\yy)$, $\e=\pm$, are the correlation functions of the two quadratic
observables $O^\e_\xx= \sum_{j=0,1}\s_{\xx}\s_{\xx+\bfe_j}+
\e\sum_{j=0,1}\s'_{\xx}\s'_{\xx+\bfe_j}$ (which are called {\it energy}, if
$\e=+$, and {\it crossover}, if $\e=-$, in the AT model, while the names are
exchanged in the 8V model), in \cite{M} it has been also proved that the large
distance decay of $G^\e(\xx-\yy)$ is faster than any power of $\x^{-1}
|\xx-\yy|$, with correlation length
\[
\x \sim  |T-T_c|^{-\n} \;, \quad\hbox{as\ } T\to T_c
\]
while at $T=T_c$, the decay of $G^\e(\xx-\yy)$ is power law:
\[
G^\e(\xx-\yy)\sim  |\xx-\yy|^{-2x_\e} \;.
\]

In the Ashkin-Teller model with $J\not= J'$, it has been proved in \cite{GM}
that there are two critical temperatures, $T_{c,1}$ and $T_{c,2}$, such that
\be
\label{h1} C_v\sim -\D^{-\a}\log [\D^{-2} |T-T_{c,1}|\cdot|T-T_{c,2}|]
\ee
where $2\D^2=(T-T_{c,1})^2+(T-T_{c,2})^2$ (the index $\a$ in \pref{h1} is
the same as in \pref{h10} ). While in the isotropic AT the logarithmic
singularity of $C_v$ is turned by the interaction in a power law, in the
anisotropic AT $C_v$ has still a logarithmic singularity; however,
$T_{1,c}-T_{2,c}$ scales with a {\it transition index} $x_T=1+O(\l)$ in the
isotropic limit:
\be
|T_{1,c}-T_{2,c}|\sim |J-J'|^{x_T}
\ee
The existence of $x_T$ was overlooked in the literature. The indices $x_+, x_-,
\n, \a, x_T$ are expressed by expansions which are {\it convergent} for $\l$
small enough. Hence, the indices can be computed in principle with arbitrary
precision by an explicit computation of the first orders and a rigorous bound
for the rest; moreover, in this way one can prove that the indices depend on
$\l$ and on all details of the model. On the other hand, the complexity of the
expansions makes essentially impossible to prove the universal relations
directly from them.

Another important class of models whose correlations can be
analyzed by similar methods are models of interacting fermions on
a 1D lattice or quantum spin chains; they are all described by the
Hamiltonian $H=$
\bea\label{12} &&-{1\over 2}\sum_{x=1}^{L-1} [a^+_{x}
a^-_{x+1}+a_{x+1}^+ a^-_{x}] -u [ a^+_{x}
a^+_{x+1}+a_{x+1}^-a^-_{x}]\\
&&+h\sum_{x=1}^L (\r_x-{1\over 2}) +\l \sum_{1\le x,y\le L}
v(x-y)(\r_x-{1\over 2})(\r_y-{1\over 2})\nn \eea
where $a^\pm_x$ are the fermion creation or annihilation operators and
$\r_x=a^+_x a^-_x$. By using the Jordan-Wigner transformation, the Hamiltonian
of the {\it Heisenberg quantum spin chains} can be written in this way, if
$J_1+J_2=2$, $u=(J_1-J_2)/2$ and $J_3=-\l$; in particular, if
$v(x-y)=\d_{|x-y|,1}/2$ and $h=0$, we have the {\it $XYZ$ model}. Let us define
$\xx=(x,x_0)$, $O_\xx=e^{H x_0} O_x e^{-H x_0}$ and, if
$A=O_{\xx_1}...O_{\xx_n}$, $\la A\ra=\lim_{L\to\io}{{\rm Tr}e^{-\b H} {\bf
T}(A)\over {\rm Tr} e^{-\b H}}$, $\bf{T}$ being the time order product. If
$u=0$, it was shown in \cite{BM} that, for $\l$ small enough, if $T$ denotes
the truncated expectation, $\la S^{(3)}_\xx S^{(3)}_{\bf 0}\ra_T\sim$
\be\label{6}
\cos(2p_F x) {1+O(\l)\over 2\p^2[x^2+(v_s x_0)^2]^{x_+}}+ {1+O(\l)\over
2\p^2[x^2+(v_s x_0)^2]}
\ee
where $S^{(3)}_\xx=\r_\xx- 1/2$, $p_F=\cos^{-1}(h+\l)+O(\l)$ is the Fermi
momentum (if $h=0$, $p_F=\pi/2$ by symmetry ) and $v_s=\sin p_F +O(\l)$ is the
Fermi (or sound) velocity, which is modified by the interaction, since,
contrary to the previous Ising case, there is no symmetry between space and
time. Finally $x_+$ is a critical index, expressed by a convergent expansion;
it depends on all details of the model, as it is apparent from the explicit
computation of its first order contribution, which gives $x_+=  1- a_1
\l+O(\l^2)$, with $a_1=[\hat v(0)- \hat v(2 p_F)]/ (\pi \sin p_F)$. When
$J_1\not=J_2$, that is $u\not=0$, $\la S^{(3)}_\xx S^{(3)}_{\bf 0}\ra_T$ decays
exponentially, with correlation length $\x\sim |J_1-J_2|^{-\bar\n}$, with
$\bar\n= 1+ a_1 \l+O(\l^2)$, $a_1$ being the same constant as before. If
$J_1=J_2$, $\la a^-_\xx a^+_\yy\ra_T \sim |\xx-\yy|^{-1-\h}$, $\h=O(\l^2)>0$,
and the correlation of the Cooper pair operator $\r^{c}_\xx=a^+_\xx
a^+_{\xx'}+a^-_\xx a^-_{\xx'}$, $\xx'=(x+1,x_0)$, decays as $|\xx-\yy|^{-2
x_-}$, $x_-=1+a_1\l+O(\l^2)$.

If $u=0$ and $j_\xx= (2i v_F)^{-1} [a^+_{\xx+1}a^-_{\xx}-
a^+_{\xx}a^-_{\xx+(1,0)}]$, the following WI, for $\kk,\kk+\pp$ close to
$\pp^\o_F \=(\o p_F,0)$, $\o=\pm$, are true:
\bea\label{WI} \label{wwi} &&-i p_0 \la \hat\r_\pp\hat a^+_{\kk}
\hat a^-_{\kk+\pp}\ra+\o p \tilde v_J \la \hat j_\pp\hat
a^+_{\kk}\hat a^-_{\kk+\pp}\ra \sim
B G\nn\\
&&-i p_0 \la \hat j_\pp\hat a^+_{\kk}\hat a^-_{\kk+\pp}\ra+\o p
\tilde v_N \la \hat \r_\pp\hat a^+_{\kk}\hat a^-_{\kk+\pp}\ra\sim
\bar B G \eea
with $G\equiv G(\kk,\kk+\pp)=[\la\hat a^+_{\kk}\hat a^-_{\kk}\ra-\la\hat
a^+_{\kk+\pp}\hat a^-_{\kk+\pp}\ra]$, $B=1$, $\bar B=1+O(\l)$ and $\tilde v_J,
\tilde v_N=v_s(1+O(\l))$; in particular $\tilde v_N / \tilde v_J=1+2
a_1\l+O(\l^2)$, with $a_1$ the constant defined above, after \pref{6}. When
$\l=0$, the continuity equations for $\r_\xx$ and $j_\xx$ imply WI similar to
\pref{WI} with $\bar B=1$ and $\tilde v_N=\tilde v_J=v_s$; the interaction has
the effect that the normalization $\bar B$ is not $1$ ( $[H, \r_x]=0$ but
$[H,j_x]\not =0$) and two different velocities, the charge $\tilde v_J$ and the
current velocity $\tilde v_N$, appear. The presence of the lattice, which
breaks the Lorentz symmetry of the continuum limit, causes the presence of
three distinct velocities, $\tilde v_N$, $\tilde v_J$, $v_s$. Finally, we
recall that the {\it susceptibility} is defined as $\k=\lim_{p\to
0}\hat\O(0,p)$, where $\hat\O(p_0,p)$ is the Fourier transform of $\la
S^{(3)}_\xx S^{(3)}_{\bf 0}\ra_T$; $\k\r^{-2}$ is the {\it compressibility} if
$\r$ is the fermionic density. Our results are contained in the following
Theorem.

{\bf Theorem} {\it Given the models with hamiltonian \pref{11}, \pref{12}, at
small coupling all the indices defined above can be uniquely expressed in terms
of one of them:
\bea
\label{xxx} &&x_- = x_+^{-1} \virg \a=2(1- x_+)(2-x_+)^{-1}
\\
\label{xxxx}&& \n^{-1} = 2-x_+ \virg \bar\n^{-1} = 2-x_+^{-1}\\
&&2\eta=x_+ + x_+^{-1}-2\label{xxxx2}\\
&&x_{T}=(2-x_+)(2-x_+^{-1})^{-1}\label{n}
\eea
Moreover, in the model \pref{12} the velocities appearing in \pref{WI} verify
$\tilde v_N \tilde v_J=v_s^2$ and $\tilde v_J=\sin p_F$, while the
susceptibility $\k$ verifies
\be
\k=x_+ (\p v_s)^{-1}\label{x11}\;.
\ee}
\pref{n} is a new relation for the Ashkin-Teller model, never proposed before;
the first relation in \pref{xxx} was conjectured in \cite{K} and \pref{xxxx},
\pref{xxxx2} in \cite{KW,LP}. \pref{x11} is part of the Haldane {\it Luttinger
liquid} conjecture \cite{H} for fermionic systems or quantum spin chains.
Some of the above relations were checked in certain solvable case: the second
of \pref{xxx}, which is equivalent, by using the first of \pref{xxxx}, to the
{\it hyper-scaling relation} $2\n=2-\a$, in the case of the Eight Vertex model
\cite{Ba}; \pref{xxxx2}, \pref{x11} in the case of the Luttinger model
\cite{Gi}; \pref{x11} in the XYZ spin chain \cite{H}. The above theorem
provides the first proof of the validity of such relations for generic non
solvable models.
Note that, in the notation of \cite{H}, $v_N\equiv (\pi\k)^{-1}$ should not be
confused with $\tilde v_N$ appearing in the WI \pref{WI}; they are coinciding
{\it only} in the special case of the Luttinger model. Therefore $\tilde
v_N=v_N$ is an example of relation true in the Luttinger model but not in the
presence of a lattice.

{\it Outline of the proof.} The technical details are long and will appear
elsewhere \cite{BFM,BM}; here we just outline the proof. The partition function
and some of the correlations of the spin model \pref{11} can be exactly
rewritten as sums of {\it Grassmann integrals} describing $d=1+1$ Dirac
fermions on a lattice and with quartic non local (but short ranged)
interactions, by using the classical representation of the Ising model in terms
of Grassmann integrals \cite{HS}. The Grassmann variables are written as
$\psi_\kk=\sum_{h=-\io}^0 \psi^{(h)}_\kk$, with $\psi^{(h)}_\kk$ living at
momentum scale $\kk=O(2^h)$. After the integration of the fields
$\psi^{(0)},\ldots,\psi^{(h+1)}$, the partition function can be written
\cite{M} as
\be\label{cc}
\int P_{Z_h,\m_h}(d\psi^{(\le h)})e^{V^{(h)}(\sqrt{Z_h}\psi^{(\le h)})}
\ee
where $P_{Z_h,\m_h}(d\psi^{(\le h)})$ is the {\it Gaussian Grassmann
integration} with propagator $g^{(\le h)}(\kk)=\frac{\chi_h(\kk')}{Z_h}\times$
\be
\begin{pmatrix}-i\sin k_0 +\sin k+\m_{++}& -\m_h-\m_{-+}\cr
-\m_h -\m_{+-}&-i\sin k_0-\sin k_1+\m_{--}\end{pmatrix}^{-1}\label{ppp}
\ee
where $\chi_h(\kk)$ is a smooth compact support function
nonvanishing only for $|\kk|\le 2^h$, $Z_h$  and $\m_h$ are the
effective wave function renormalization and the effective mass,
$\m_{\pm\pm}$ are $O(\kk^2)$ and non vanishing at $\kk=(\pm
\pi,\pm\pi)$ (there is no fermion doubling problem); moreover,
$\VV^{(h)}=\l_h\sum_\xx \psi^+_{\xx,+} \psi^-_{\xx,+}
\psi^+_{\xx,-} \psi^-_{\xx,-} +R_h$, where $\l_h$ is the effective
coupling and $R_h$ is a sum of {\it irrelevant terms}, represented
as space-time integrals of field monomials, multiplied by kernels
which are {\it analytic functions} of $\l_k$, $k>h$. Analyticity
is a very non trivial property, obtained via {\it tree expansions}
\cite{BGPS} and exploiting anticommutativity properties of Grassmann
variables, via {\it Gram inequality} for determinants (which takes
into account compensations among different graphs of different
signs at a given order). It is important to stress that \pref{cc}
is {\it exact}, in the sense that the irrelevant terms and the
lattice are fully kept into account (in standard RG applications
they are instead neglected). The effective coupling $\l_h$
converges, as $h\to-\io$, to a function $\l_{-\io}$ (analytic
function of $\l$), thanks to the asymptotic vanishing of the beta
function, which is a consequence of Ward Identities. A similar
analysis can be repeated in the case of the fermionic model
\pref{12}, the main (but trivial) difference being that \pref{ppp}
is replaced by a similar expression, taking into account that
$x_0$ is a continuous variable; such asymmetry has the effect
that, contrary to what happens in the spin case, the velocity is
{\it renormalized} by the interaction. In order to exploit the
asymptotic symmetries of the model, it is convenient to introduce
the following Grassmann integral
\be\label{cc1}
\int P^{th}_{Z}(d\psi^{(\le N)})e^{V^{(N)}(\sqrt{Z_N}\psi^{(\le N)})}
\ee
where, if $\psi = (\psi_+,\psi_-)$ and $\bar\psi=\psi^+\g_0$ are Euclidean
$d=1+1$ spinors, $P^{th}_{Z}(d\psi^{(\le N)})$ is the fermionic gaussian
integration with propagator $ g^{(\le N)}(\kk)=\chi_N(\kk) (\g_\m
\kk_\m)^{-1}$, and $V^{(N)}(\psi^{(\le N)})=\int d\xx d\yy v(\xx-\yy) j_\m(\xx)
j_\m(\yy)$, with $j_\m(\xx)= \bar\psi_\xx \g_\m \psi_\xx$ and $v(\xx-\yy)$ a
short range symmetric interaction. A multiscale integration is now necessary
also in the ultraviolet region to perform the limit $N\to\io$, while in the
integration of the infrared scales an expression similar to \pref{cc} is found;
the effective coupling is denoted by $\tilde\l_h$. The crucial point is that it
is possible to choose, by a  fixed point argument, the values of
$\tilde\l_{\io}$ (fixed $c=\sin p_F$ in the model \pref{11}, while $c=v_s$ in
the model \pref{12}) so that $\l_{-\io}=\tilde\l_{-\io}$. This implies that
{\it the critical exponents of the two models are the same}, because the
exponents are expressed by series in $\tilde\l_{-\io}/c$ with universal
coefficients. Of course $\tilde\l_\io$ is a convergent series in $\l$ depending
on all details of the models \pref{11} or \pref{12}. On the other hand, the
continuum Grassmann integral \pref{cc1} verifies extra Lorentz and Gauge
symmetries, implying exact Ward Identities when the ultraviolet cut-off is
removed; by the transformation $\psi\to e^{i\a_\xx}\psi_\xx$ one finds
$-i\pp_\m \la j_{\m,\pp}\psi_{\kk}\bar\psi_{\kk+\pp}\ra_{th}=$
\be
\la \psi_{\kk}\bar\psi_{\kk}\ra_{th}- \la
\psi_{\kk+\pp}\bar\psi^-_{\kk+\pp}\ra_{th} +\D_N(\kk,\pp)\label{wixx}
\ee
where $\D_N=\la \d j_{\pp}\psi_{\kk}\bar\psi_{\kk+\pp}\ra_{th}$, with $\d
j_{\pp}=\int d\kk [(\chi^{-1}_N(\kk+\pp)-1)(\g_\m \kk_\m + \g_\m \pp_\m)
-(\chi^{-1}_N(\kk)-1) \g_\m \kk_\m]\bar\psi_{\kk}\psi_{\kk+\pp}$; an analogous
expression is obtained for the axial current $\bar\psi\g_\m\g_5\psi$.
By a multiscale
analysis it can be proved that $\lim_{N\to\io} \D_N(\kk,\pp)=$
\be -i\t\hat v(\pp) \pp_\m \la j_{\m,\pp}\psi_{\kk,\o}
\bar\psi_{\kk+\pp,\o}\ra_{th}\virg \t=\tilde\l_{\io}/
(4\pi c)
\ee
A similar expression holds for the chiral WI; the fact that $\D_N(\kk,\pp)$ is
not vanishing in the limit $N\to\io$ is a manifestation of a {\it quantum
anomaly}. The anomaly coefficient $\t$ is {\it linear} in $\tilde\l_\io$; this
is the non-perturbative analogue of the {\it anomaly non renormalization} in
QED in 4D. Such crucial property depends on our assumption about the
interaction in \pref{cc1}; it would not be true, for instance, if we replace
$v(\xx-\yy)$ with a delta function \cite{BFM1}. By combining the WI with the
Schwinger-Dyson equations, one gets some equations for the correlations, from
which the indices can be computed as functions of $\t$. One can find, for
example, that $x_+= (1- \t) ( 1+\t)^{-1}$, $x_-= (1+ \t) ( 1-\t)^{-1}$, so that
$x_+ x_-=1$; the other relations  follow by similar
arguments. Note that the indices we consider have a simple expression in terms
of $\tilde\l_\io$, but $\tilde\l_\io$ is of course rather complex and model
dependent as a function of $\l$.

A similar RG analysis can be repeated for the model \pref{12}; it turns out
that the vertex functions in the first line of \pref{WI} are asymptotically
coinciding with $ Z^{(3)}\la j^0_\pp\hat
\psi^+_{\kk,\o}\hat\psi^-_{\kk+\pp,\o}\ra_{th}$ and $i \tilde Z^{(3)}\la
j^1_\pp\hat \psi^+_{\kk,\o}\hat\psi^-_{\kk+\pp,\o}\ra_{th}$, with $\tilde
Z^{(3)}/ Z^{(3)}=1+a_1\l+O(\l^2)$; therefore, by using the WI for the model
\pref{cc1}, we derive \pref{WI} with $B=Z^{(3)} Z^{-1} (1-\t)^{-1}=1$, $\bar
B=\tilde Z^{(3)} Z^{-1} (1+\t)^{-1}$, $\tilde v_N=v_s Z^{(3)}/ \tilde Z^{(3)}$,
$\tilde v_J=v_s \tilde Z^{(3)}/ Z^{(3)}$; on the other hand, the equations of
motion related to the lattice Hamiltonian impose the constraints $v_J=\sin p_F$
and $B=1$. Finally a WI for the density correlation can be also derived; if
$D_\o(\pp)=-i p_0+\o v_s p$ and $\hat\O(\pp)= \la \hat\r_\pp \hat\r_\pp\ra=$,
we get
\[ \hat\O(\pp)=
{1\over 4\pi v_s Z^2}{(Z^{(3)})^2\over 1-\t^2} \left[2-{D_-(\pp)\over
D_+(\pp)}-{D_+(\pp)\over D_-(\pp)} \right]
\]
which implies \pref{x11}, by using $\k=\lim_{p\to 0} \hat\O(0,p)$.

In conclusion, we have established for the first time the validity of a number
of universal relations among critical exponents and other quantities in a
wide class of generally non solvable lattice models. They are true in special
continuum solvable models and we have proven that the lattice symmetry breaking
effects produce different velocities in the model \pref{12} and change the
critical exponents, but do not destroy the validity of several universal
relations (on the other hand, not all the relations valid in the solvable
models are generically true, like the relation $\tilde v_N=v_N$ or the
relations for the dynamic exponents \cite{IG}). Some of the universal relations
are used for the analysis of experiments in carbon nanotubes or spin chains,
but we believe that their interest goes much beyond this, as they provide one
of the very cases in which the {\it universality principle}, a general belief
in Statistical Physics and beyond, can be rigorously verified. Extensions of
our methods will hopefully allow to prove universal relations in an even wider
class of models, as well as other relations among spin or dynamic exponents.


\begin{thebibliography}{999999}

\bibitem{Au} O.M. Auslaender {\it et al.}, Phys. Rev. Lett. {\bf 84}, 1764
    (2000); M. Bockrath {\it et al.}, Nature {\bf 397}, 598 (1999);
    H. Ishiii {\it et al.}, Nature {\bf 426}, 540 (2003); Z. Yao {\it et al},
    Nature {\bf 401}, 273 (1999).

\bibitem{BP} Per Bak {\it et al.}, Phys. Rev. Lett. {\bf 54}, 1539 (1985);
N.C. Bartelt, T.L. Einstein, Phys. Rev. B {\bf 40}, 10759 (1989).

\bibitem{Ba} R.J. Baxter, Academic Press, London, (1989).

\bibitem{BFM1} G. Benfatto, P. Falco, V. Mastropietro, Comm. Math. Phys.
    {\bf 273}, 67 (2007).

\bibitem{BFM} G. Benfatto, P. Falco, V. Mastropietro, Comm. Math. Phys., to
    appear (2009).

\bibitem{BGPS} G. Benfatto, G. Gallavotti, A. Procacci, B. Scoppola, Comm.
    Math. Phys. {\bf 160}, 93 (1994).

\bibitem{BM} G. Benfatto, V. Mastropietro, arXiv:0907.2837.


\bibitem{Gi} T. Giamarchi, Oxford University Press, Oxford (2004).

\bibitem{GM} A. Giuliani, V. Mastropietro, Phys. Rev. Lett. {\bf 93},
    190603--07 (2004); Comm. Math. Phys. {\bf 256}, 681 (2005).

\bibitem{K} L.P. Kadanoff, Phys. Rev. Lett. {\bf 39}, 903 (1977).

\bibitem{KB} L.P. Kadanoff, A.C. Brown, Ann. Phys. {\bf 121}, 318 (1979).

\bibitem{KW} L.P. Kadanoff, F. Wegner, Phys. Rev. B
    {\bf 4}, 3989 (1971).

\bibitem{H} F.D.M. Haldane, Phys. Rev. Lett. {\bf 45}, 1358 (1980).

\bibitem{HS} C.A. Hurst, H.S. Green, J. Chem. Phys. {\bf 33}, 1059 (1960).

\bibitem{IG} A. Imambekov, L.I. Glazman, Phys. Rev. Lett. {\bf 100},
    206805 (2008); Science {\bf 323}, 228 (2009); Phys.
    Rev. Lett. {102}, 126405 (2009)

\bibitem{L} B. Lake {\it et al.}, Nature materials {\bf 4}, 329 (2005).

\bibitem{LP} A. Luther, I. Peschel, Phys. Rev.B {\bf 12}, 3908 (1975).

\bibitem{M} V. Mastropietro, Comm. Math. Phys. {\bf 244}, 595 (2004).


\bibitem{N} M.P.M. den Nijs, Phys. Rev. B {\bf 23}, 6111 (1981).

\bibitem{PB} A.M.M. Pruisken, A.C. Brown, Phys. Rev. B {\bf 23}, 1459
    (1981); A.M.M. Pruisken, L.P. Kadanoff, Phys. Rev. B {\bf 22} 5154
    (1980).

\bibitem{SML} T.D. Schultz, D.C. Mattis, E.H. Lieb, Rev. Mod. Phys. {\bf
    36}, 856 (1964).

\bibitem{PS} T. Spencer, Physica A {\bf 279}, 250 (2000); H. Pinson,
    T. Spencer, unpublished.

\bibitem{S} H. Spohn, Phys. Rev. E {\bf 60}, 6411 (1999).

\bibitem{ZZ} A.B. Zamolodchikov,  Al.B. Zamolodchikov, Soviet Scientific
    Reviews A {\bf 10}, 269 (1989).

\end{thebibliography}
\end{document}